\documentclass[12pt]{article}
\begin{document}
\title{Negative Energy Solutions and Symmetries}
\author{B.G. Sidharth\\
International Institute for Applicable Mathematics \& Information Sciences\\
Hyderabad (India) \& Udine (Italy)\\
B.M. Birla Science Centre, Adarsh Nagar, Hyderabad - 500 063
(India)}
\date{}
\maketitle
\begin{abstract}
We revisit the negative energy solutions of the Dirac equation,
which become relevant at very high energies and study several
symmetries which follow therefrom. The consequences are briefly
examined.
\end{abstract}
\vspace{5 mm}
\begin{flushleft}
03.65.-W; 03.65.Pm
\end{flushleft}
\section{Introduction}
It is well known that in relativistic Quantum Mechanics, we
encounter negative energy solutions, be it for the Dirac equation or
for the Klein-Gordon equation. Such negative energy solutions have
no counterpart, indeed interpretation in non relativistic or
classical theory. For the Klein-Gordon (K-G) equation, this could be
attributed to the second time derivative, which leads to an extra
degree of freedom. Pauli and Weiskoef interpreted the negative
energy solutions in the context of Quantum Field Theory, but what is
less well known is that these negative energy solutions of the
Klein-Gordon equation were successfully interpreted thereafter by
Feshbach and Villars \cite{feshbach} in the context of the usual
single particle theory.\\
It could have been expected that these difficulties would be
bypassed in the Dirac theory which restores the single time
derivative -- but here too the negative energies surfaced, because
ultimately it was the same energy momentum dispersion relation that
was invoked. Dirac then had to take recourse to the negative energy
sea and the hole theory to overcome the difficulty \cite{dirac}.
Interestingly a different explanation was given by the author
several years ago in the context of Quantum Mechanical Kerr Newman
Black Holes \cite{bgsijpap}. We will now study the negative energy
solutions for both the Dirac and Klein-Gordon equations and examine
some symmetries and also their consequences.
\section{The Negative Energy Solutions}
Let us write the Dirac wave function as
\begin{equation}
\psi = \left(\begin{array}{ll} \phi \\
\chi\end{array}\right),\label{3.27}
\end{equation}
where $\phi$ and $\chi$ are each two spinors. As is well known
\cite{bd}, we can then deduce
$$\imath \hbar (\partial \phi / \partial t) = c \tau \cdot (p -
e/cA) \chi + (mc^2+e\phi)\phi ,$$
\begin{equation}
\imath \hbar (\partial \chi / \partial t) = c \tau \cdot (p - e/cA)
\phi + (-mc^2+e\phi)\chi.\label{3.29}
\end{equation}
We recapitulate that at low energies $\chi$ is small and $\phi$
dominates, whereas it is the reverse at high energies. We also note
that while sensible wave packets can be formed with the positive
energy solutions alone, in general we require both signs of energy
for a localized particle. In fact the Compton wavelength is the
minimum extension, below which both positive and negative solutions
will have to be considered. Well outside the Compton wavelennth, we
can continue with the usual positive energy description. More
formally the positive energy solutions alone do not form a complete
set of eigen functions of the Hamiltonion.\\
The following symmetry can be seen from (\ref{3.29}) (with $e = 0$,
or the absence of an external electromagnetic field for simplicity):
\begin{equation}
t \to -t, \phi \to - \chi\label{3}
\end{equation}
We must remember that we are dealing with intervals at the Compton
scale, so that the negative energy solutions are relevant. So the
time reversal given in (\ref{3}) is at the Compton scale.\\
We next observe that such a $(t,-t)$ behaviour in this microscopic
interval has been described in detail in terms of a double Weiner
process. Furthermore this can be used in the context of two state
systems to go over from the non relativistic Schrodinger theory to
the relativistic theory (Cf.\cite{bgsfpl162003,uof,tduniv} for
details). To see this briefly, we first define a complete set of
base states by the subscript $\imath \quad \mbox{and}\quad U
(t_2,t_1)$ the time elapse operator that denotes the passage of time
between instants $t_1$ and $t_2$, $t_2$ greater than $t_1$. We
denote by, $C_\imath (t) \equiv < \imath |\psi (t)
>$, the amplitude for the state $|\psi (t) >$ to be in the state $|
\imath >$ at time $t$. We have \cite{bgsijpap,cu,bgscsfqfst}
$$< \imath |U|j > \equiv U_{\imath j}, U_{\imath j}(t + \Delta t,t) \equiv
\delta_{\imath j} - \frac{\imath}{\hbar} H_{\imath j}(t)\Delta t.$$
We can now deduce from the super position of states principle that,
\begin{equation}
C_\imath (t + \Delta t) = \sum_{j} [\delta_{\imath j} -
\frac{\imath} {\hbar} H_{\imath j}(t)\Delta t]C_j (t)\label{2xe}
\end{equation}
and finally, in the limit,
\begin{equation}
\imath \hbar \frac{dC_\imath (t)}{dt} = \sum_{j} H_{\imath
j}(t)C_j(t)\label{2fe}
\end{equation}
where the matrix $H_{\imath j}(t)$ is identified with the
Hamiltonian operator. We have argued earlier at length that
(\ref{2fe}) leads to the Schrodinger equation
\cite{bgsijpap,bgscsfqfst}. In the above we have taken the usual
unidirectional time to deduce the non relativistic Schrodinger
equation. If however we consider a Weiner process in (\ref{2xe})
that is, allow $t$ to fluctuate between $(t - \Delta t, t + \Delta
t)$, (to which we will return shortly), then we will have to
consider instead of (\ref{2fe})
\begin{equation}
C_\imath (t - \Delta t) - C_\imath (t + \Delta t) = \sum_{j}
\left[\delta_{\imath j} - \frac{\imath}{\hbar} H_{\imath
j}(t)\right] C_j (t)\label{2ge}
\end{equation}
Equation (\ref{2ge}) in the limit can be seen to lead to the
relativistic Klein-Gordon equation rather than the Schrodinger
equation with the second time derivative \cite{bgscsfqfst,tduniv}.
In other words the symmetry in (\ref{3}) is in-built at the Compton
scale in the relativistic description, be it for the Klein-Gordon
equation or the Dirac equation, and Zitterbewegung is a
manifestation of this (Cf.\cite{feshbach,bgsijtp}).\\
We can push these considerations further. We have already seen the
symmetry given in (\ref{3}): In case of a charged particle, in
addition, $e \to -e$ and vice versa (with complexification). This
apart it suggests that the coordinate $\vec{x}$, as it were splits
into the coordinate $\vec{x}_1$ and $\vec{x}_2$ which mimic the wave
function in (\ref{3.27}) at low and high energies, in the sense that
the former dominates at low energies while the latter dominates at
high energies, following the wave function as in (\ref{3.27}). The
fact that these go into each other following (\ref{3}) as $t \to -t$
can be explained in terms of the development of a two Weiner process
see briefly above (Cf.\cite{tduniv}). Let us elaborate.\\
In this case there are two derivatives, one for the usual forward
time and another for a backward time given by
\begin{equation}
\frac{d_+}{dt} x (t) = {\bf b_+} \, , \, \frac{d_-}{dt} x(t) = {\bf
b_-}\label{2ex1}
\end{equation}
where we are considering for the simplicity, a single dimension $x$.
This leads to the Fokker-Planck equations
$$
\partial \rho / \partial t + div (\rho {\bf b_+}) = V \Delta \rho
,$$
\begin{equation}
\partial \rho / \partial t + div (\rho {\bf b_-}) = - U \Delta
\rho\label{2ex2}
\end{equation}
defining
\begin{equation}
V = \frac{{\bf b_+ + b_-}}{2} \quad ; U = \frac{{\bf b_+ - b_-}}{2}
\label{2ex3}
\end{equation}
We get on addition and subtraction of the equations in (\ref{2ex2})
the equations
\begin{equation}
\partial \rho / \partial t + div (\rho V) = 0\label{2ex4}
\end{equation}
\begin{equation}
U = \nu \nabla ln\rho\label{2ex5}
\end{equation}
It must be mentioned that $V$ and $U$ are the statistical averages
of the respective velocities and their differences. We can then
introduce the definitions
\begin{equation}
V = 2 \nu \nabla S\label{2ex6}
\end{equation}
\begin{equation}
V - \imath U = -2 \imath \nu \nabla (l n \psi)\label{2ex7}
\end{equation}
We will not pursue this line of thought here but refer the reader to
Smolin \cite{smolin} for further details. We now observe that the
complex velocity in (\ref{2ex7}) can be described in terms of a
positive or uni directional time $t$ only, but a complex coordinate
\begin{equation}
x \to x + \imath x'\label{2De9d}
\end{equation}
To see this let us rewrite (\ref{2ex3}) as
\begin{equation}
\frac{dX_r}{dt} = V, \quad \frac{dX_\imath}{dt} = U,\label{2De10d}
\end{equation}
where we have introduced a complex coordinate $X$ with real and
imaginary parts $X_r$ and $X_\imath$, while at the same time using
derivatives with respect
to time as in conventional theory.\\
We can now see from (\ref{2ex3}) and (\ref{2De10d}) that
\begin{equation}
W = \frac{d}{dt} (X_r - \imath X_\imath )\label{2De11d}
\end{equation}
That is we can alternatively use derivatives with respect to the
usual uni directional time derivative
to introduce the complex coordinate (\ref{2De9d}) (Cf.ref.\cite{bgsfpl162003}.\\
Let us now generalize (\ref{2De9d}), which we have taken in one
dimension for simplicity, to three dimensions. Then as discovered by
Hamilton, we end up with not three but four dimensions,
$$(1, \imath) \to (I, \tau),$$
where $I$ is the unit $2 \times 2$ matrix and $\tau$s are the Pauli
matrices. We get the special relativistic \index{Lorentz}Lorentz
invariant metric at the same time. (In this sense, as noted by Sachs
\cite{sachsgr}, Hamilton would have hit upon \index{Special
Relativity}Special Relativity, if he had identified the new fourth
coordinate
with time).\\
That is,\\
\begin{equation}
x + \imath y \to Ix_1 + \imath x_2 + jx_3 + kx_4,\label{Aa}
\end{equation}
where $(\imath ,j,k)$ momentarily represent the \index{Pauli}Pauli
matrices; and, further,
\begin{equation}
x^2_1 + x^2_2 + x^2_3 - x^2_4\label{B}
\end{equation}
is invariant.\\
While the usual \index{Minkowski}Minkowski four vector transforms as
the basis of the four dimensional representation of the
\index{Poincare}Poincare group, the two dimensional representation
of the same group, given by the right hand side of (\ref{Aa}) in
terms of \index{Pauli}Pauli matrices, obeys the quaternionic algebra
of the second rank
\index{spin}spinors (Cf.Ref.\cite{bgsfpl162003,shirokov,sachsgr} for details).\\
In fact one representation of the two dimensional form of the
\index{quarternion}quarternion basis elements is the set of
\index{Pauli}Pauli matrices above. Thus a
\index{quarternion}quarternion may be expressed in the form
$$Q = -\imath \tau_\mu x^\mu = \tau_0x^4 - \imath \tau_1 x^1 - \imath \tau_2 x^2 -
\imath \tau_3 x^3 = (\tau_0 x^4 + \imath \vec \tau \cdot \vec r)$$
This can also be written as
$$Q = -\imath \left(\begin{array}{ll}
\imath x^4 + x^3 \quad x^1-\imath x^2\\
x^1 + \imath x^2 \quad \imath x^4 - x^3
\end{array}\right).$$
As can be seen from the above, there is a one to one correspondence
between a \index{Minkowski}Minkowski four-vector and $Q$. The
invariant is now given by
$Q\bar Q$, where $\bar Q$ is the complex conjugate of $Q$.\\
In this description we would have from (\ref{Aa}), returning to the
usual notation,
\begin{equation}
[x^\imath \tau^\imath , x^j \tau^j] \propto \epsilon_{\imath jk}
\tau^k \ne 0\label{y}
\end{equation}
In other words, as (\ref{y}) shows, the coordinates no longer follow
a commutative geometry. It is quite remarkable that the
noncommutative geometry (\ref{y}) has been studied by the author in
some detail (Cf.\cite{tduniv}), though from the point of view of
Snyder's minimum fundamental length, which he introduced to overcome
divergence difficulties in Quantum Field Theory. Indeed we are
essentially in the same situation, because as we have seen, for our
positive energy description of the universe, there is the minimum
Compton wavelength cut off for a meaningful description
\cite{bgsextn,schweber,newtonwigner}.\\
Proceeding further we could think along the lines of $SU (2)$ and
consider the transformation \cite{taylor}
\begin{equation}
\psi (x) \to exp [\frac{1}{2} \imath g \tau \cdot \omega (x)] \psi
(x).\label{4.2}
\end{equation}
This leads as is well known to a covariant derivative
\begin{equation}
D_\lambda \equiv \partial_\lambda - \frac{1}{2} \imath g \tau \cdot
W_\lambda,\label{4.3a}
\end{equation}
remembering that $\omega$ in this theory is infinitessimal. We are
thus lead to vector Bosons $W_\lambda$ and an interaction like the
strong interaction, described by
\begin{equation}
W_\lambda \to W_\lambda + \partial_\lambda \omega - g \omega \Lambda
W_\lambda.\label{4.4}
\end{equation}
However, we are this time dealing, not with iso spin, but between
positive and negative energy states as in (\ref{3.27}). Also we must
bear in mind that this non-electromagnetic force between particles
and anti particles would be valid only within the Compton time,
inside this Compton scale Quantum Mechanical "bridge" \cite{report}.\\
These considerations are also valid for the Klein-Gordon equation in
the two component notation developed by Feshbach and Villars
\cite{feshbach,uheb}. There too, we get equations like (\ref{3.29}).
We would like to re-emphasize that our usual description in terms of
positive energy solutions is valid above the Compton scale.
\section{A Further Symmetry}
As we consider both signs of the energy, we denote the expectation
of an operator by the equation (Cf. also ref.\cite{feshbach})
\begin{equation}
\int \psi^* \tau_3 \Omega \psi d^3 x\label{A}
\end{equation}
where $\tau_3$ is the usual Pauli matrix is given by
\begin{equation}
\psi^* = (\phi , \chi) = \tau_3 = \left(\begin{array}{ll} 1 \quad 0\\
0 \quad -1\end{array}\right)\label{A1}
\end{equation}
We use (\ref{A}) for the observable: $\Omega = x^\imath x_\imath$.
Then we can easily see the following. Let us first consider
(\ref{A}) for the two cases: First the negative energy component
$\chi$ is vanishingly small, as in our usual description and second
where the negative energy component dominates and $\phi$ is
vanishingly small, that is for the very high energy case. Then we
can easily verify that
$$\Omega \to - \Omega$$
This has the following consequence. The Minkowski invariant
\begin{equation}
x^\mu x_\mu\label{x1}
\end{equation}
of the Lorentz group goes over
to the invariant of the four dimensional rotation group
\begin{equation}
x^2_0 + x^\imath x_\imath\label{y1}
\end{equation}
for negative energies and vice versa.\\
We could expect that the Foldy-Wothuysen transformation goes over to
a Lorentz transformation in the negative energy realm. A simple way
of seeing this is as follows: The Foldy-Wothuysen transformation is
given by
$$S = e^{\beta \vec{\alpha} \cdot \vec{p} \Theta}$$
\begin{equation}
tan \, 2|p| \Theta = \frac{|p|}{m}\label{c1}
\end{equation}
while the Lorentz transformation is described by
$$S = -e^{-\imath \vec{\alpha} \cdot \vec{p} (\mu)}$$
\begin{equation}
tan \, h \mu |p| = \frac{pc}{E + mc^2}\label{d1}
\end{equation}
(Cf.ref.\cite{bd}). Comparison of (\ref{x1}) and (\ref{y1}) show
that effectively $x^j \to \imath x^j$ or $p_j \to \imath p_j$. Under
this transformation (\ref{c1}) and (\ref{d1}) get interchanged.
\section{Remarks}
i) As mentioned, the above considerations for the Dirac equation all
apply for the positive and negative energy solutions of the Klein-Gordon equation
(Cf.\cite{uheb}.\\
ii) We make the following remark about the negative and positive
energy solutions of the Dirac equation. We consider for simplicity
the free particle solutions \cite{bd}. The solutions are of the type
\begin{equation} \psi = \psi_A + \psi_S\label{e25}
\end{equation}
where
$$
\psi_A =   e^{\frac{\imath}{\hbar} Et} \ \left(\begin{array}{l}
                                          0 \\ 0 \\ 1 \\ 0
                             \end{array}\right) \mbox{ or } Ü e^{\frac{\imath}{\hbar} Et}
                          \ \   \left(\begin{array}{l}
                                 0 \\ 0 \\ 0 \\ 1
                              \end{array}\right) \mbox{ and }
$$
\begin{equation}
\label{e26}
\end{equation}
$$
\psi_S =  e^{-\frac{\imath}{\hbar} Et} \ \left(\begin{array}{l}
                                 1 \\ 0 \\ 0 \\ 0
                               \end{array}\right) \mbox{ or } e^{-\frac{\imath}{\hbar} Et}
                            \ \   \left(\begin{array}{l}
                                 0 \\ 1 \\ 0 \\ 0
                               \end{array}\right)
$$
denote respectively the negative energy and positive energy
solutions. From (\ref{e25}) the probability of finding the particle
in a small volume about a given point is given by
\begin{equation}
| \psi_A + \psi_S|^2 = |\psi_A|^2 + |\psi_S|^2 + (\psi_A \psi_S^* +
\psi_S \psi_A^*)\label{e27}
\end{equation}
Equations (\ref{e26}) and (\ref{e27}) show that the negative energy
and positive energy solutions form a coherent Hilbert space and so
the possibility of transition to negative energy states exists. This
difficulty however can be overcome by the well known Hole theory
which uses the Pauli exclusion
principle, and is described in many standard books on Quantum Mechanics.\\
However the last or interference term on the right side of
(\ref{e27}) is like the zitterbewegung term. When we remember that
we really have to consider averages over space time intervals of the
order of $\hbar/mc$ and $\hbar/mc^2$ as Dirac himself pointed out
(Cf.\cite{dirac}), this term disappears and effectively the negative
energy solutions and positive energy solutions stand decoupled in
what is now
the physical universe.\\
A more precise way of looking at this is\cite{schweber} that as is
well known, for the homogeneous Lorentz group, $\frac{p_0}{|p_0|}$
commutes with all operators and yet it is not a multiple of the
identity as one would expect according to Schur's lemma: The
operator has the eigen values $\pm 1$ corresponding to positive and
negative energy solutions. This is a super selection principle or
"spin" referred to in (\ref{4.2}) pointing to the two incoherent or
decoupled Hilbert spaces or universes \cite{r31} now represented by
states $\psi_A$ and $\psi_S$ which have been decoupled owing to the
averaging over the Compton wavelength space- time intervals which
eliminates the interference term in (\ref{e27}). But all this refers
to energies such that our length scale is greater than the Compton
wavelength.\\
Thus once again we see that outside the Compton wavelength region we
recover the usual physics.\\
iii) It is worth recapitulating that we have identified the negative
energy solutions with anti particles and via the mechanism described
by (\ref{e27}), that is based on the fact that physical measurements
are time averages over intervals of the order of the Compton scale.
We conclude that the anti particles are very short lived, because
outside the Compton wavelength that is in our physical world, we are
in the manifold of positive solutions. Further these considerations
also show (Cf.refs.\cite{uheb,bgsmod}) that there is an asymmetry
between particles and anti particles. Indeed this prediction has
since been suggested through experiment: Firstly there is the
observed neutrino and anti neutrino asymmetry that violates CP,
observed in the MiniBooNE experiment at Fermilab recently.
Specifically the oscillation patterns for the neutrino and anti
neutrino appear to be different with a confidence level of about
$99.7\%$ This in fact corroborates an earlier LSND experiment report
at the Los Alamos National Laboratory in 1990, but since not taken
seriously because it appeared too sensational.\\
The other CP violation has been found in the so called B factories
at SLAC, US and KEK Lab in Japan. This collaboration has calculated
that the parameter associated with CP violation -- '$sin 2 \beta$'
-- is $0.74 \pm 0.07$, compared with its earlier estimate of $0.99
\pm 0.14$. The increased accuracy stems from the larger number of
decay events observed this time -- $88$ million in total. The BELLE
collaboration puts the value of $siine 2 \beta$ -- which they call
$sin e 2 \psi_1$ -- at $0.79 \pm 0.10$
\cite{aubert2005,aubert2004}.\\
The new estimates established beyond doubt that CP violation exists.

\end{document}